# AI-Enhanced Sensemaking: Exploring the Design of a Generative AI-Based Assistant to Support Genetic Professionals

Angela Mastrianni[*]

College of Computing and Informatics, Drexel University, am4673@drexel.edu

Hope Twede

Microsoft Research, hope.twede@microsoft.com

Aleksandra Sarcevic

College of Computing and Informatics, Drexel University, as3653@drexel.edu

Jeremiah Wander

Microsoft Research, miah@microsoft.com

Christina Austin-Tse

Broad Institute of MIT and Harvard, caustint@broadinstitute.org

Scott Saponas

Microsoft Research, ssaponas@microsoft.com

Heidi Rehm

Broad Institute of MIT and Harvard, hrehm@broadinstitute.org

Ashley Mae Conard[±]

Microsoft Research, ashleyconard@microsoft.com

Amanda K. Hall[±]

Microsoft Research, amanda.hall@microsoft.com

Generative AI has the potential to transform knowledge work, but further research is needed to understand how knowledge workers envision using and interacting with generative AI. We investigate the development of generative AI tools to support domain experts in knowledge work, examining task delegation and the design of human-AI interactions. Our research focused on designing a generative AI assistant to aid genetic professionals in analyzing whole genome sequences (WGS) and other clinical data for rare disease diagnosis. Through interviews with 17 genetics professionals, we identified current challenges in WGS analysis. We then conducted co-design sessions with six genetics professionals to determine tasks that could be supported by an AI assistant and considerations for designing interactions with the AI assistant. From our findings, we identified sensemaking as both a current challenge in WGS analysis and a process

---

[*] Contact Author (This work was done during the author's internship at Microsoft Research), ± Equal Contribution

that could be supported by AI. We contribute an understanding of how domain experts envision interacting with generative AI in their knowledge work, a detailed empirical study of WGS analysis, and three design considerations for using generative AI to support domain experts in sensemaking during knowledge work.

## 1 INTRODUCTION

Recent technological advances in generative AI have the potential to transform knowledge work [35, 72, 90]. *Knowledge work* is the process of applying existing knowledge to solve problems, sharing knowledge with others, and creating new knowledge [41]. *Sensemaking*, a critical part of knowledge work, involves collecting, synthesizing, and building representations of information [66, 71, 87]. Unlike prior advancements in automation which mainly impacted routine or manual tasks, generative AI models can perform sensemaking tasks typically associated with knowledge work [15, 35, 72]. In this study, we examine the design of generative AI tools to support domain experts in their knowledge work, situating our research in the context of whole genome sequencing (WGS) analysis for rare disease diagnosis. During WGS analysis, an individual's genome is compared to reference genomes to detect variants that may be causing their disease. To analyze these variants, genetic professionals engage in sensemaking to build models of information known about genes and variants by finding, synthesizing, and interpreting information from several knowledge sources, a time- and effort-intensive process. Because new information about genes and variants is continuously being discovered, analysts must stay abreast of new scientific findings and engage in sensemaking to understand how these findings could impact previously unsolved cases. As fewer than half of individuals with rare diseases receive a diagnosis from the initial analysis of their WGS data, cases may be reanalyzed at later points to identify and apply any new information about gene-disease relationships [83]. Several challenges impede WGS analysis reaching a higher diagnostic yield, including the limited time of available genetics professionals to address the rising number of unsolved cases and limited computational support [2]. As a result, the development of AI tools has been identified as a key opportunity to increase diagnostic yield [18].

Although generative AI has the potential to support knowledge work, concerns have arisen that generative AI could intensify current social issues, such as disinformation, biased decision-making, deskilling, and job replacement [24, 90]. To mitigate these issues, a *human-centered approach* can be used to develop generative AI tools. In this approach, intelligent systems are designed with a focus on the people using the system and an understanding that they are part of a larger sociotechnical system [6, 70, 91]. We follow a human-centered approach in this study, focusing on three research questions (RQ): (RQ1) What are the current challenges and needs of genetic professionals performing WGS analysis? (RQ2) What tasks could an AI assistant support? and (RQ3) How do genetic professionals envision interacting with the AI assistant? To answer these questions, our study contains two phases: (1) needs elicitation to answer RQ1 and (2) design ideation to answer RQ2 and RQ3. In the first study phase (*needs elicitation*), we interviewed 17 genetics professionals to better understand their workflows, tools, and challenges. These genetics professionals included analysts directly involved in interpreting WGS data, as well as other roles participating in whole genome sequencing. In the second phase (*design ideation*), we ideated on the design of a generative AI assistant to support WGS analysis, focusing on the potential uses of *large language models* (LLMs). LLMs are built from large amounts of text data and can perform a variety of tasks, synthesizing knowledge across information sources and creating human-like content [7, 8, 84]. We conducted co-design sessions, including a group workshop and individual follow-up sessions, with six genetic professionals. In the group workshop, participants brainstormed tasks that an AI assistant could support and sketched potential interactions with the AI assistant in their workflows. We then drew on the workshop findings to develop a prototype of an AI assistant, which was used as a design probe in the follow-up design walk-through sessions.

During the needs elicitation phase of the study, we identified three main challenges in WGS analysis: (1) aggregating and synthesizing information about a gene and variant, (2) sharing findings with other genetic professionals, and (3) prioritizing cases for reanalysis. Genetic professionals highlighted the time-consuming nature of gathering and synthesizing information about genes and variants from different data sources. Other genetic professionals may have insights into certain genes and variants, but sharing and interpreting information with others for collaborative sensemaking requires significant time and effort (if done at all). Although new scientific findings could impact unsolved cases through reanalysis, prioritizing cases for reanalysis based on new scientific findings was challenging due to the number of unsolved cases and limited time of genetic professionals. In design ideation phase, participants envisioned and prioritized two potential tasks of an AI assistant: (1) flagging cases for reanalysis based on new scientific findings, and (2) aggregating and synthesizing key information about genes and variants from scientific publications. We also identified two themes in their feedback about these features. First, participants highlighted the need to balance selective and comprehensive evidence at different points in their workflows. Second, participants envisioned collaboratively interpreting, editing, and verifying artifacts generated by the AI assistant with other genetic professionals.

Our findings highlight how genetic professionals envisioned and prioritized the use of generative AI to support their sensemaking. The AI assistant co-designed with our participants could support sensemaking by helping to aggregate, synthesize, and apply the growing information about genes and variants to rare disease cases. Additionally, sharing the AI-generated artifacts between analysts could facilitate collaborative sensemaking and reduce the work required to assess the trustworthiness of artifacts. Our paper makes three main contributions to the human-computer interaction (HCI) community. First, we identify how genetic professionals envision generative AI being used to support their knowledge work. Second, through a detailed empirical study of WGS analysis, we provide insight into how genetic professionals find, make sense of, and share information. Third, we propose three design considerations for aiding individual and collaborative sensemaking with generative AI: (1) facilitating distributed sensemaking, (2) supporting initial sensemaking and re-sensemaking, and (3) combining multiple modalities of evidence. Although our study is situated in the context of WGS analysis, our findings and contributions could provide insight into other areas where domain experts perform sensemaking tasks as part of their knowledge work.

## 2 BACKGROUND: WHOLE GENOME SEQUENCING FOR RARE DISEASE DIAGNOSIS

Collectively, rare diseases have been estimated to affect up to half a billion people around the world [59]. For individuals with a rare disease, finding a diagnosis can take multiple years and involve specialist consultations, laboratory tests, imaging studies, and invasive procedures [76]. The aim of whole genome sequencing in rare disease diagnosis is to identify the variant or variants in an individual's genome that are causing their disease. A DNA sequence is extracted from the individual's sample (e.g., their blood or saliva) and aligned to reference sequences to detect differences (genetic variants) [2, 56]. Variant analysts then analyze the sequence by filtering, prioritizing, and classifying the variants (Figure 1). This process is computer-mediated with bioinformatics tools developed to support analysis and interpretation [20, 61]. Typically, bioinformatics pipelines will be used to annotate the variants with several different types of information, such details about the associated gene and the position of the variant within the gene [58] (Figure 1.1). This information can then be leveraged to filter the millions of variants to a smaller subset that may be further evaluated for disease causality. For example, if the analyst has access to medical information and samples from the individual's parents, they may apply specific filters based on inheritance rules (Figure 1.2). Even after applying standard filters, hundreds of variants can be returned in a typical case [2]. Analysts then need to triage the list of variants to identify the variants that may be pathogenic and causal for the individual's disease (Figure 1.3). For example, an analyst may prioritize a variant if the individual's presentation of their disease (known as their phenotype) matches diseases known to be associated with the gene in which that variant resides. When triaging and interpreting variants, analysts make sense of a wide variety of information sources to better understand the gene and variant. The information is aggregated from many sources including scientific publications, reference databases (e.g., DECIPHER [27]), and manually curated databases (e.g., Online Mendelian Inheritance in Man (OMIM) [30], Gene Curation Coalition (GenCC) [21], and ClinVar [44]). DECIPHER, OMIM, and GenCC provide information about gene-disease associations, while ClinVar aggregates user-submitted variant pathogenicity classifications. In some instances, analysts may pursue additional laboratory testing to validate or refine variant call and interpretation, such as confirming a variant with low quality metrics [2] (Figure 1.4). After triaging variants, analysts will classify the pathogenicity of the prioritized variants. Many research and clinical groups follow guidelines from the American College of Medical Genetics and Genomics (ACMG) when classifying and reporting variants [69] (Figure 1.5). These guidelines detail the process and evidence that should be used when classifying a variant along a spectrum of categories from "benign" to "pathogenic". The entire WGS process takes around 3-12 weeks, with an average of ~7.3 hours per case actively spent on variant analysis [2][1,2,3].

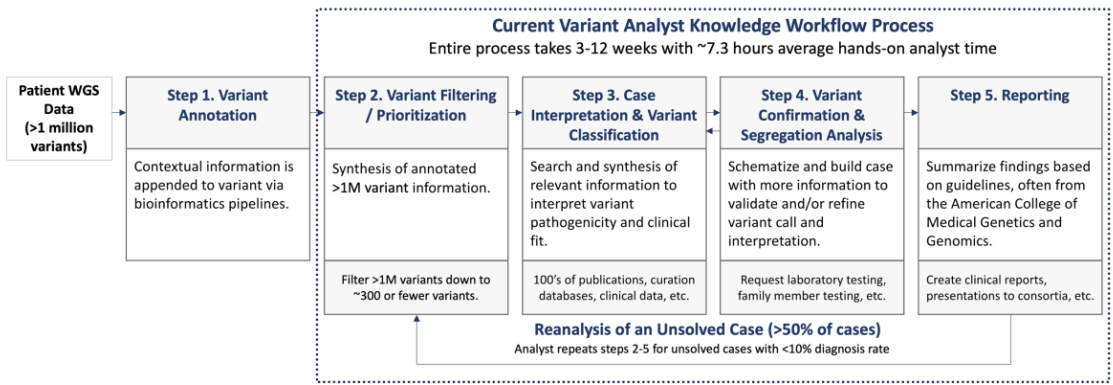

Figure 1: Variant analyst knowledge workflow process in the context of variant analysis. Steps (1-5) within the model of the current workflow for the analysis of an individual's WGS (whole genome sequencing) data are derived from [2]. The dotted boundary box represents steps within the workflow which are part of analyst knowledge work.

Because fewer than half of cases are solved with the initial analysis, periodic reanalysis of WGS data can increase the diagnostic yield of WGS [18]. During periodic reanalysis, analysts repeat the same steps as their initial analysis, with the goal of understanding if there is any new information that influences their interpretation of an individual's WGS data (Figure 1.6). Factors that can lead to a new diagnosis in reanalysis include new scientific findings (e.g., a publication about a novel gene discovery), improvements to bioinformatics pipelines, and updates to phenotype information. For example, over 1,000 papers are published each year with new disease descriptions or new gene-disease links that can potentially

---

[1] Prevention Genetics WGS turnaround time: https://www.preventiongenetics.com/pgnome/ Last Visited: 9/16/2024
[2] Baylor Genetics WGS turnaround time: https://catalog.baylorgenetics.com/details/1810 Last Visited: 9/16/2024
[3] Variantyx WGS turnaround time: https://www.variantyx.com/products-services/rare-disorder-genetics/comprehensive-analyses/genomic-unity-whole-genome-analysis/ Last Visited: 9/16/2024

lead to a diagnosis in a previously unsolved case [22]. However, only about 10% of cases that are periodically reanalyzed are solved, meaning the remainder would require another reanalysis to apply any new information that could produce a diagnosis. This aggregation of unsolved reanalyzed cases, plus new unsolved cases, results in an ever-increasing backlog of cases qualifying for reanalysis. Although reanalysis may lead to additional solved cases, implementation of reanalysis is organization dependent. For example, some WGS providers offer limited periodic reanalysis as a service included in the original test order while others only offer reanalysis as a separate test ordered by the physician[4,5].

## 3 RELATED WORK

Next, we discuss related work on designing human-centered AI systems and using AI to support knowledge workers in sensemaking.

### 3.1 Designing Human-Centered AI Systems with/for Domain Experts

Around 87% of AI systems do not reach deployment, because these systems are often solving the wrong problems [88]. Understanding the types and forms of AI support desired by different communities of practice has therefore been an open research area [90]. Appropriately matching the capabilities of a technology with the user tasks that could be most beneficially impacted by the technology has historically been a challenging problem [5]. This problem is further accentuated when designing AI-based technologies because AI models can generate many possible outputs and the capabilities of these models can be unclear [93]. Fundamental considerations in designing AI-based systems include determining the tasks supported by AI, the appropriate level of AI involvement in those tasks, and the design of interactions between users and AI [52, 82].

Participatory and community-based research approaches can be used to directly engage domain experts in envisioning and designing AI systems [17, 90]. These types of approaches are useful in designing human-centered AI systems, but several challenges exist [12, 93]. First, domain experts can face issues envisioning and critiquing potential uses of AI without appropriately understanding its capabilities and limitations [12, 93]. To mitigate this challenge, participants can be provided with knowledge about AI capabilities before starting design ideation [1, 96]. A second challenge in using participatory and community-based research approaches is encouraging participants to imagine novel human-AI interactions. Participants often fixate on existing interaction forms, such as chatbots, recommendations, and alerts [95]. During participatory design approaches, participants can be prompted to consider different degrees of AI delegation, such as: no AI, machine-in-the-loop, human-in-the-loop, and AI only [52]. In the machine-in-the-loop level of delegation, a person is responsible for performing a task, with assistance from AI when appropriate. In the human-in-the-loop level, the AI is largely performing a task, with confirmation from a person when necessary.

Recent studies have used participatory approaches to engage domain experts in designing AI systems, identifying tasks in their workflows that could be supported by AI and factors influencing their desired interactions with AI [80, 94, 95, 97]. These domain experts included speech pathologists [80], radiologists [94, 97], and intensive care clinicians [95], with participants envisioning AI systems supporting tasks such as documentation, care coordination, and billing [80, 94]. When envisioning potential interactions with AI systems, domain experts often desired interactions that would facilitate

---

[4]Prevention Genetics WGS reanalysis policy: https://www.preventiongenetics.com/ClinicalTesting/re-analysis-policy Last Visited: 09/16/2024
[5]Baylor Genetics WGS reanalysis policy: https://www.baylorgenetics.com/news/test-launch-whole-genome-sequencing-reanalysis Last Visited: 09/16/2024

explainability, feedback, and trust between the system and users [95, 97]. For example, users wanted the ability to view explanations from an AI system to help in determining their trust in the system's output [97].

In this study, we focus on designing AI tools for genetic professionals performing WGS analysis for rare disease diagnosis. Several automated and AI-based tools have been developed to support WGS by prioritizing variants for review [4, 19, 45, 50, 57]. Studies of these tools have historically focused on evaluating technical performance, not on how these tools may be used by genetic professionals or incorporated within their workflows. Further research from a human-centered design perspective can provide insight into the challenges faced by analysts during WGS analysis and the opportunities for AI support. Prior HCI studies on genome sequencing and rare disease diagnosis have mainly focused on designing technologies to aid individuals in exploring their genomic data [78] and supporting individuals with rare diseases [54, 55, 60]. In this study, we consider a novel group within the rare disease diagnosis process (genetic professionals) and propose new technological solutions to support their work. In addition to supporting rare disease diagnosis, studying the work of genetic professionals provides insights into how knowledge workers engage in highly complex tasks and make sense of vast amounts of new and existing knowledge. From our study, we find that our participants envisioned and prioritized using generative AI to support their sensemaking during WGS analysis.

### 3.2 Developing Tools to Support Knowledge Workers in Sensemaking

Knowledge workers aim to understand a body of knowledge and use this information to solve problems and generate new information for their organization [42]. A critical part of knowledge work involves sensemaking. The sensemaking process consists of two interconnected activities: (1) *foraging* (searching, filtering, and synthesizing information) and (2) *sensemaking* (building, refining, and presenting models of information) [66]. The process of sensemaking can differ depending on the domain. For example, prior research has focused on supporting researchers in making sense of scientific literature and identifying new research areas [13, 39, 40, 62, 99]. Clinicians, however, may have different needs when making sense of scientific literature. Clinicians often focus on assessing the applicability of papers to their patients and may stop searching for literature once they have gathered enough evidence to justify a decision [92]. In collaborative environments, such as healthcare settings, sensemaking is often a social and interactive activity that occurs between people [64]. Even when people do not directly collaborate, individuals may be able to leverage the artifacts and work of another person to improve their own sensemaking (known as distributed sensemaking) [25]. However, users may decide not to use the sensemaking artifacts of others if the work required to interpret those artifacts is too high. Conversely, creating sensemaking artifacts in a manner that is useful to other users may be too time- and effort-intensive for the current user [46]. When deciding to reuse knowledge from another person's sensemaking process, people often consider the trustworthiness of the other person, the context of their sensemaking (e.g., the goals motivating their sensemaking), and the thoroughness of their sensemaking [48].

Prior work has developed tools to support sensemaking in a range of contexts [13, 29, 39, 40, 43, 62, 68, 99]. A subset of tools help users in creating artifacts that could not only support their sensemaking, but be shared with others to support their sensemaking as well [46, 86, 98]. For example, Unakite [46] assists users in building comparison tables to better understand the tradeoffs between different options when searching for information online. However, these tools still require user effort to capture, synthesize, and organize information. To reduce this burden, automated approaches can be used. For example, Crystalline [47] leverages natural language processing (NLP) to automatically extract relevant information from webpages and organize this information into comparison tables to support decision making. With the recent advances in LLMs, these models can be even more effective at extracting and synthesizing information, avoiding some of the errors found with earlier NLP approaches [49]. This is largely due to LLMs being trained on incredibly vast amounts of diverse text data, resulting in comprehensive understanding of written language capable of flexible reasoning over many kinds of text structures [7, 8, 84].

Because of the impressive capabilities of LLMs, recent work has explored how these models can be used to support sensemaking [37, 40, 49, 63, 81]. For example, ChoiceMates [63] gives users the ability to converse with multiple LLM-powered agents for sensemaking support when making unfamiliar decisions. Many popular and mainstream LLM tools (e.g., ChatGPT[6], BingChat[7], Gemini[8]) facilitate interactions between the user and LLM through a conversational interface. However, conversational interfaces may not support users in tracking information shared throughout the conversation, which can limit their sensemaking [67, 81]. As a result, recent studies have investigated alternative interaction mechanisms to support sensemaking [37, 40, 49, 81]. For example, one study found that users preferred interactive diagrams over conversational interfaces when interacting with an LLM to gather and interpret information [81]. Through our study, we make two contributions to research on generative AI support of sensemaking. First, we examine how generative AI may be used to support the sensemaking of genetic professionals. Unlike other studies that focused on supporting everyday sensemaking [37, 49, 63, 81], we investigate how generative AI can support domain experts in sensemaking. Second, we explore alternative interaction mechanisms in addition to conversational interfaces and propose design considerations for AI-enhanced sensemaking.

## 4 STUDY METHODS

Our study had two phases: (1) needs elicitation and (2) design ideation (). In the first phase, we conducted interviews with genetic professionals to understand the challenges faced during WGS analysis (RQ1 – needs elicitation). In the second phase, we conducted co-design sessions to design an AI assistant to support WGS analysis (RQ2 – AI tasks, RQ3 – AI interaction). This study was approved by our Institutional Review Board.

### 4.1 Primary Research Site and their Platform for Sequencing Analysis (*seqr*)

We have an ongoing collaboration with our primary research site, the Broad Institute, an academic research institute in the United States. The Broad Institute primarily performs genome sequencing for internal research studies, as well as studies led by external research collaborators. The Broad Institute developed and currently uses an open-source, web-based tool for rare disease case analysis and project management (*seqr [61]* ). To facilitate genome sequencing analysis, *seqr* supports variant filtration, annotation, and presumed causal variant identification. *Seqr* has a project page, where analysts can view all cases in a project and filter for the cases assigned to the analyst. After selecting a case, *seqr* opens the family page, which contains medical information about the individual with the rare disease and their family. Analysts can also record notes about the case on this page. Analysts can then use a search feature to investigate variants in the individual's genome. This search feature includes a list of predefined searches which can be customized as needed. The searches each return a list of gene-variant pairs. Each gene-variant pair is annotated with the information needed to interpret its pathogenicity and causality, such as the type of variant, information submitted about the variant in ClinVar, and any disease associations for the gene found in OMIM. Analysts can also use a dedicated section for taking notes about the gene-variant pair and can flag the gene-variant pair for further review or exclusion.

---

[6] https://chat.openai.com/ Last Visited: 05/01/2024
[7] https://www.bing.com/chat Last Visited: 05/01/2024
[8] https://gemini.google.com/ Last Visited: 05/01/2024

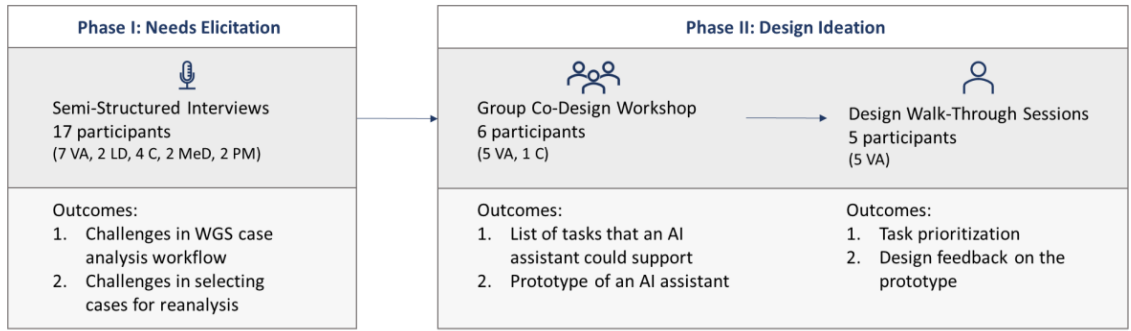

Figure 2: An overview of the study phases. Eighteen genetic professionals participated across the two study phases. Genetic professional participant roles in WGS (whole genome sequencing) analysis: variant analyst (VA), laboratory director (LD), clinician (C), methods developer (MeD), program manager (PM)

### 4.2 Participant Recruitment

Our participants were recruited from five different roles determined to be relevant to the task of WGS analysis in rare disease analysts: variant analyst (VA), laboratory director (LD), clinician (C), methods developer (MeD), and program manager (PM) (Table 1). Variant analysts, laboratory directors, and clinicians are directly involved in interpreting WGS data and many of the tasks performed by these three roles may overlap. The primary role of variant analysts, also known as variant scientists, genomic analysts, or clinical analysts, is analyzing and interpreting variants. They also have interpretation-related responsibilities, such as medical record review and case management. Laboratory directors review and approve the work of variant analysts, and conduct variant interpretation and project management. Clinicians may have various roles within the rare disease case analysis ecosystem including patient-facing roles of ordering genome analysis and relaying relevant findings to individuals with rare diseases or lending their clinical experience to molecular laboratories or research programs. Although methods developers and program managers may not be directly involved in interpreting WGS data, we included participants from these roles in the interviews because their work affects the analysis process. Methods developers (MD) create computational toolkits and platforms used in the analysis. Program managers (PM) are higher-level administrators who determine the processes of the lab, clinic, or research program. Because the genetics professional network is relatively small in rare disease, participant information is presented in aggregate to ensure anonymity (Table 1).

We recruited 17 genetic professionals (seven variant analysts, two laboratory directors, two clinicians, two method developers, and two program managers) to participate in Phase I interviews. We first sent a call for participation email to our primary research site and recruited fourteen individuals. We also used snowballing sampling and recruited three additional participants from external sites: a variant analyst and a program manager outside of the United States, and a methods developer in the United States.

For the Phase II design ideation sessions, we recruited six genetic professionals (five variant analysts and one clinician). Of these six participants, five had also participated in the earlier interviews. Recruitment for the Phase II was constrained to individuals actively performing case interpretation, primarily variant analysts. To ground the design thinking within a specific tool, we limited participants to those who had familiarity with *seqr*, the platform planned for use in our prototype

Table 1: An overview of the study participants (n=18). 17 genetic professionals participated in Phase I and six genetics professionals participated in Phase II. Five of the Phase II participants had participated in Phase I.

| Role | # of Participants | Participant Credentials | Participant Experience Range |
| --- | --- | --- | --- |
| Variant Analyst (VA) | 8 | BS (1), MS/CGC (3), PhD (4) | Between 1-5 years to more than 10 years |
| Laboratory Director (LD) | 2 | PhD/ABMGG(1), PhD/HCLD/CGC (1) | Between 5-10 years to more than 10 years |
| Clinician (C) | 4 | MS/LCGC (2), MD/PhD (1), MD (1) | Between 1-5 years to between 5-10 years |
| Methods Developer (MeD) | 2 | BS (1), PhD (1) | Less than 1 year to between 5-10 years |
| Program Manager (PM) | 2 | MS/LCGC (1), PhD/FFSc (1) | Between 5-10 years to more than 10 years |

BS=Bachelor of Science, MS=Master of Science, MD=Doctor of Medicine, PhD=Doctor of Philosophy, CGC=Clinical Genetic Counselor, LCGC=Licensed Clinical Genetic Counselor, ABMGG=American Board of Medical Genetics and Genomics, HCLD=Laboratory Director of High Complexity Testing, FFSc=Fellow of the Faculty of Science development. We sent a call for participation email to eligible participants from the Phase I interviews, followed by snowballing sampling for additional participants.

### 4.3 Phase I: Interviews (RQ1 – Needs Elicitation)

*4. 3. 1 Protocol.* We conducted all interviews over a video-conferencing platform (Figure 2). The interview structure and questions were developed and curated by our research team in collaboration with non-participant genetic professionals from our primary research site. In the first interview section, we asked about the participant's day-to-day responsibilities, challenges faced, and tools used. In the second section, we asked about the data used in their WGS analysis and the opportunities and limitations of the data. In the third section, we discussed their process for reanalyzing cases, asking questions about the initialization "triggers" for reanalysis, the reanalysis workflow, and challenges in reanalysis. The semi-structured interviews were conducted by a member of the research team with prior experience in genomic analysis. We recorded audio and video of the interviews after obtaining consent from each participant.

*4. 3. 1 Data Analysis.* We performed an inductive, qualitative content analysis [16] of the interview transcripts, focusing on the (1) tools and data used in WGS analysis, (2) reanalysis initialization triggers, and (3) challenges in the initial analysis and reanalysis of WGS data. Transcripts were automatically generated from the interview recordings and then verified for accuracy. An HCI researcher on the team iteratively open-coded the transcripts and connected codes to develop a preliminary set of themes. The themes were discussed and further refined by our multidisciplinary research team, which included researchers with backgrounds in HCI, genomics, and computer science.

### 4.4 Phase II: Co-Design Sessions (RQ2 – AI Tasks, RQ3 – AI Interaction)

We conducted co-design sessions with variant analysts at our primary research site. Co-design sessions directly engage users in the design process, empowering them to brainstorm and prioritize new technologies [73]. We conducted two types of co-design sessions: (1) a group co-design workshop and (2) individual design walk-through sessions.

*4. 4. 1 Group Co-Design Workshop: Protocol.* Six participants participated in the group workshop (Figure 2). Four participants joined the workshop in person and two joined remotely. Four researchers facilitated the workshop (one remotely and three in person). Two researchers have a background in HCI, one has a background in computer science and computational biology, and one has experience in WGS analysis. Drawing on prior work [1, 96], we blended user-centered and technology-centered approaches in conducting the workshop. The half-day workshop had four sections: (1) a discussion of the interview findings from Phase 1, (2) an overview of the recent advances in AI, (3) a brainstorming activity on the tasks supported by an AI assistant, and (4) a sketching activity on the design of interactions with an AI assistant. In the first section, we reviewed the findings from interviews to both validate findings with participants and prompt thinking about the challenges faced in WGS analysis. Next, we asked participants to discuss their experiences in using generative AI tools (e.g., ChatGPT, BingChat). A researcher on the team with experience in developing AI tools followed by presenting an overview of AI. We provided a brief history of AI, a description of foundational models, examples of generative AI tools, and limitations of generative AI (e.g., lack of explainability, possibility of hallucinations). To broaden the design thinking of participants, we also provided examples of systems that have other interaction mechanisms besides conversational user interfaces (e.g., [89]). We concluded the presentation by listing some capabilities of AI based on design resources from prior work [96]. Next, we conducted a brainstorming activity, where we asked participants to individually write down tasks that an AI assistant could support. Participants then shared their ideas with the wider group. Finally, we asked participants to individually sketch how they imagine interacting with an AI assistant in their existing platform for analysis and reanalysis. Participants shared these sketches with the group. One participant voluntarily left the workshop before the sketching activity because they no longer wanted to participate in the session.

*4. 4. 2 Group Co-Design Workshop: Data Analysis.* We used affinity diagramming to analyze the data collected during the group co-design workshop. This method is well-suited for wide-ranging and unstructured types of data and can be used to highlight patterns between issues and areas for potential innovation [3, 53]. From the group co-design workshop, we collected three types of data: (1) a list of tasks generated during the brainstorming activity, (2) participants' sketches, and (3) notes taken by team researchers during the session. We analyzed the tasks brainstormed by participants by consolidating related tasks and iteratively grouping tasks into categories. We also analyzed participant sketches and their descriptions of the sketches to identify features and interaction mechanisms to be explored through prototyping. Drawing on these findings, we created a clickable prototype of an AI assistant within the analysis tool used at our primary research site.

*4.4.3 Individual Design Walk-Through Sessions: Protocol.* We used the prototype developed from the group co-design workshop findings as a probe to elicit discussion and design exploration in individual design walk-through sessions. We conducted the individual design walk-through sessions one week after the group workshop. Five of the six participants from the group workshop participated in a session (Figure 2). The virtual 45-minute sessions began with a prioritization activity, asking the participant to select the top three tasks that an AI assistant should support during WGS analysis. We also asked if there were any tasks that the participant thought an AI assistant should not support. We then presented the participant with the AI assistant prototype. After participants discussed their initial impressions of the AI assistant's features, we asked follow-up questions to further probe their feedback, which we used to iteratively refine the prototype. The sessions were recorded with participant consent.

*4.4.3 Individual Design Walk-Through Sessions: Data Analysis.* We analyzed the transcripts from the individual design walk-through sessions by conducting an inductive, qualitative content analysis [16]. During this analysis, we focused on understanding themes in participant feedback to the prototype. An HCI researcher on the team iteratively open-coded the transcripts and connected codes to form an initial set of themes. These themes were discussed and refined through discussions with the wider research team.

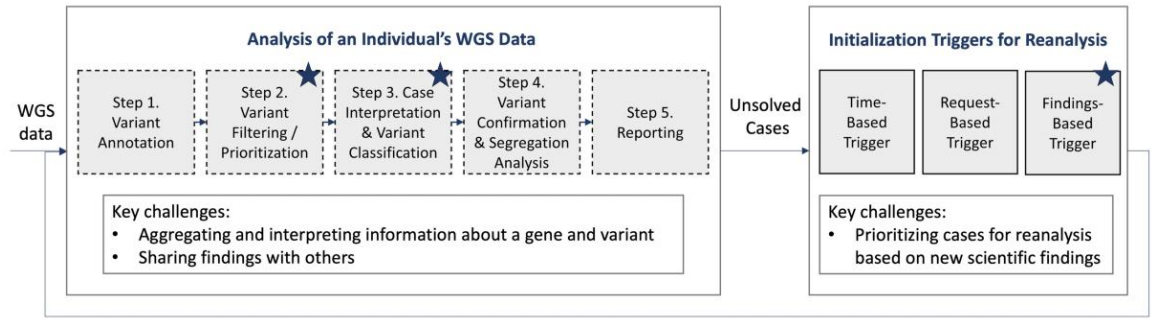

Figure 3: A model of the workflow for the initial analysis and reanalysis of WGS (whole gene sequencing) data. The workflow steps for the initial analysis (represented with dotted lines) are derived from prior work [2]. We identified initialization triggers for reanalysis from Phase I interviews. The star icons indicate the points in the workflow envisioned and prioritized by participants for AI support during Phase II individual design walk-through sessions.

## 5 PHASE I FINDINGS: USER NEEDS AND CHALLENGES IN WGS ANALYSIS (RQ1)

In response to RQ1, we identified three main challenges in WGS analysis from our interview data (Figure 3). The first two challenges occur when an analyst is performing an initial analysis or reanalysis of an individual's WGS data. The third challenge occurs when analysts select cases for reanalysis.

### 5.1 Looking within a Case: Analyzing an Individual's WGS Data

During an initial analysis or reanalysis of WGS data, analysts need to review and interpret the variants in an individual's WGS data. Analysts may share their findings in several ways, including presenting variants of interest to team members for their feedback, submitting findings to data-sharing platforms, and preparing results for clinicians. Below we discuss two challenges that we identified in this process (Figure 3).

*5.1.1 Challenge One: Aggregating and Interpreting Information about a Gene and Variant.* When reviewing and interpreting gene-variant pairs in a case, analysts need to engage in foraging and sensemaking to aggregate and synthesize information known about the gene and variant to interpret if the individual's variant could be causing their disease. Because information about genes and variants can be found in many different online sources, retrieving and synthesizing this information can be time-consuming: *"I just feel like we still have to go to so many different resources, particularly for literature searching… none of it is in one place. We're doing Google Scholar searches, we're going to HGMD [a database of papers on variants and genes [79]], we are pulling the literature from ClinVar and just concatenating and all that wastes so much time…"* [LD#1]. Analysts do not want to miss a piece of information that could change their interpretation of a case, which makes it challenging to decide when to stop investigating the information sources and move on to other variants or cases: *"I think a lot of time is also spent trying to make sure you haven't missed something"* [VA#6]. Several participants highlighted the need for computational tools to better support this process: *"Automated information gathering and then on top of that, if the information could be organized in a useful way that reflects how I would then be categorizing that collated data, that would be helpful* [VA#7]. Participants noted that reanalysis could be performed more efficiently if they had more effective tools for understanding the new information since the last analysis. When reanalyzing cases, analysts currently repeat sensemaking work done in the initial analysis to determine if there have been any changes in information about genes and variants: *"Sometimes it feels like we're repeating work that we've already done before…we have to go out and check all these different databases once again, and oftentimes there isn't new information"* [VA#1].

*5.1.2 Challenge Two: Sharing Findings with Others.* Analysts often share their findings with others to get feedback on their interpretations and identify any other information about a gene or variant. Although analysts can get feedback from other team members at group meetings, preparing presentations for the meetings can be time-consuming and tedious: *"Creating presentations can always be challenging, not because they're by nature difficult, but the tedious, repetitive nature... I feel like it's quite clunky to have to every time create a presentation from scratch because they all follow the same pattern, even in my last group"* [VA#6]. Analysts may also use data-sharing platforms to share and discuss their findings with other researchers or clinical groups outside their institution. Data-sharing platforms can facilitate connections with others who may be investigating a gene or variant, allowing analysts to obtain information that has not yet been formally published in a scientific paper. For example, when an analyst discovers a novel gene that they suspect may be related to a disease, they can submit information about the gene to the MatchMaker Exchange [65]. This platform connects the analysts to others who may be investigating that gene or have a case with a similar phenotype. As the platform has grown and become more saturated with submissions, manually following up with others interested has become more time-consuming: *"Chances are if you put a gene into MatchMaker Exchange, you're going match with at least one person and I've had in some cases, well over 50 matches... To follow up with these people individually by email... find out what their patient's presentation is, what the genotype is. All of that adds a lot of logistics time"* [VA#3]. In general, participants highlighted the need to improve data-sharing platforms so that people would be encouraged to share their findings and could easily do so in a format that is useful for others: *"Where is that sweet spot between sharing enough that somebody could analyze and reanalyze a case? And not so much that people start to skip out on it because they're like 'I just don't have time to fill this entire data model out'"* [PM#2].

## 5.2 Looking Across Cases: Selecting Cases for Reanalysis

Unsolved cases may be reanalyzed at later points in time to understand if new information about a gene or variant could be used to solve the case. We identified three types of potential initialization triggers for the reanalysis of WGS data: (1) time-based, (2) request-based, and (3) findings-based (Figure 3). With the *time-based* trigger, cases are reanalyzed because a certain amount of time has elapsed since the last analysis. With the *request-based* trigger, cases are reanalyzed because of a request from the individual, their clinician, or family. For groups that analyze cases as part of research studies, the request for reanalysis may come from their study collaborators. With the *findings-based* trigger, cases are reanalyzed because of a new scientific finding (e.g., a new scientific paper or new method) that could affect the case. Each clinical or research group performing the sequencing analysis often had policies around which of the three initialization triggers would lead to reanalysis of their cases. For example, one group at our primary research site tried to reanalyze cases every year (time-based trigger). Once reanalysis is triggered, the reanalysis workflow is similar to the initial analysis workflow. Analysts will frequently do a full reanalysis of the case, repeating the steps taken during the initial analysis. Reanalysis can take between minutes to several hours, depending on several factors, including the availability of sequences from the individual's family (which allow analysts to rule out variants based on patterns of inheritance) and the number of publications about a gene and variant. If the case has detailed notes from prior analyses, analysts can refer to those notes to reanalyze the case more quickly.

*5.2.1 Challenge Three: Prioritizing Cases for Reanalysis.* Because less than half of cases are solved with the initial WGS analysis, the backlog of unsolved cases that need to be reanalyzed keeps increasing as analysts receive and analyze new cases. With this growing backlog, participants highlighted challenges in selecting cases for reanalysis, noting a need for better ways of prioritizing cases: *"Some of these families literally have a reminder in their diary for it's been a year, have you found anything new, which is heart-breaking... but at some point, we have to prioritize, do we look at [a case]*

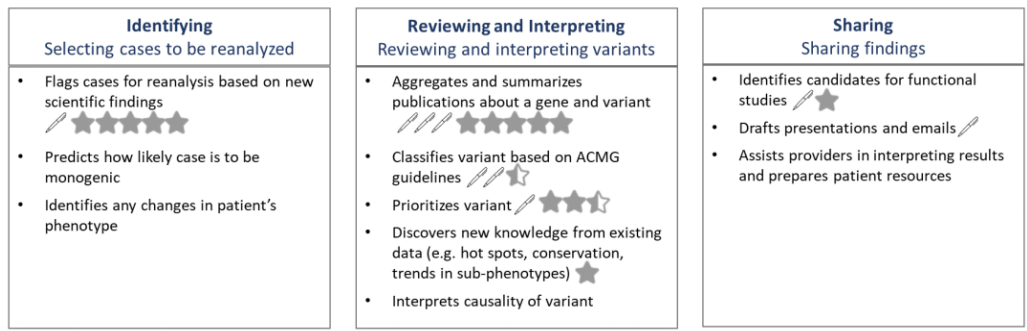

Figure 4: Tasks that could be supported by an AI assistant as brainstormed by participants during the group co-design workshop. A pencil icon indicates that a participant included this feature in their sketch of an AI assistant. Each design walkthrough participant (n=5) was asked to prioritize three tasks. A star icon indicates that a participant voted to prioritize this task in the individual follow-up sessions (e.g., five stars indicates that all participants voted to prioritize task). Half-filled stars indicate participant choosing to prioritize both tasks, though at a lower priority than their other two prioritized tasks.

*because it's going to be something interesting or because someone's asked...*" [VA#7]. For reanalysis triggered by time or request, no guarantee exists that anything has changed since the last analysis, which can be discouraging and make it unlikely that the reanalysis will be successful: *"You reanalyze every case as long as the clinician asks you to, right? That means you don't know if there's new information available, therefore the actual value proposition and the chance of solving the case at that point in time is quite low..."* [PM#1]. New evidence in the literature could help solve cases but analysts lacked a tool that would automatically identify relevant new papers, synthesize the findings, and identify potentially impacted cases: *"If this could be automated, that would be great, like this is a new intellectual disability gene that was published this week, here are the cases that have rare intronic variants in that gene... because right now, it's kind of random, like if I go to conferences, I have a list of genes to look at..."* [C#2]. However, reimbursement models can pose challenges to systems that automatically reanalyze cases. In some countries, government funding systems will only reimburse a reanalysis when it is triggered by the request of a clinician: *"In a scenario where we have an automated reanalysis system, it will undoubtedly solve many more cases. But there's no way for me to get paid for it"* [PM#1].

## 6 PHASE II FINDINGS: TASKS AND DESIGN OF AI ASSISTANT TO SUPPORT WGS ANALYSIS (RQ2, RQ3)

In response to RQ2 and RQ3, we describe findings from the design ideation phase of the study, where we focused on "designing the right thing" (RQ2 - determining tasks of an AI assistant) and "designing the thing right" (RQ3 - exploring interactions with the AI assistant) [10, 96].

### 6.1 Designing the "Right Thing": Determining Tasks of an AI Assistant (RQ2)

During the group co-design workshop, participants brainstormed tasks that could be supported by an AI assistant. After consolidating the list of tasks (Figure 4), we asked participants in the individual follow-up sessions to vote on the top three tasks that an AI assistant should support and describe the rationale behind their votes. Two tasks were prioritized by all participants: (1) flag cases for reanalysis based on new scientific findings and (2) aggregate and synthesize key information about a gene and variant from scientific publications. From the rationales for their votes, we identified two factors that were frequently considered by participants. First, participants often balanced the strengths of the AI and the limitations of analysts. They prioritized tasks that require a lot of effort or are currently not done and that they thought an AI assistant could perform well. For example, when asked why they prioritized aggregating and synthesizing publications about a gene and variant, one participant explained: *"I feel like I spend the most time digging into a gene, trying to figure out what's known about it... that's a lot of wasted analysis time ... that AI could do much faster than I can do..."* [VA#5]. Second, participants often considered the likelihood that AI support of a task would lead to external impact. For example, one participant [VA#2] discussed how they were deciding between voting on the "aggregating and synthesizing publications" and "discovering new knowledge from existing data" tasks. The participant prioritized aggregating and synthesizing publications because they thought support of this task would more likely lead to a returnable result for their patient.

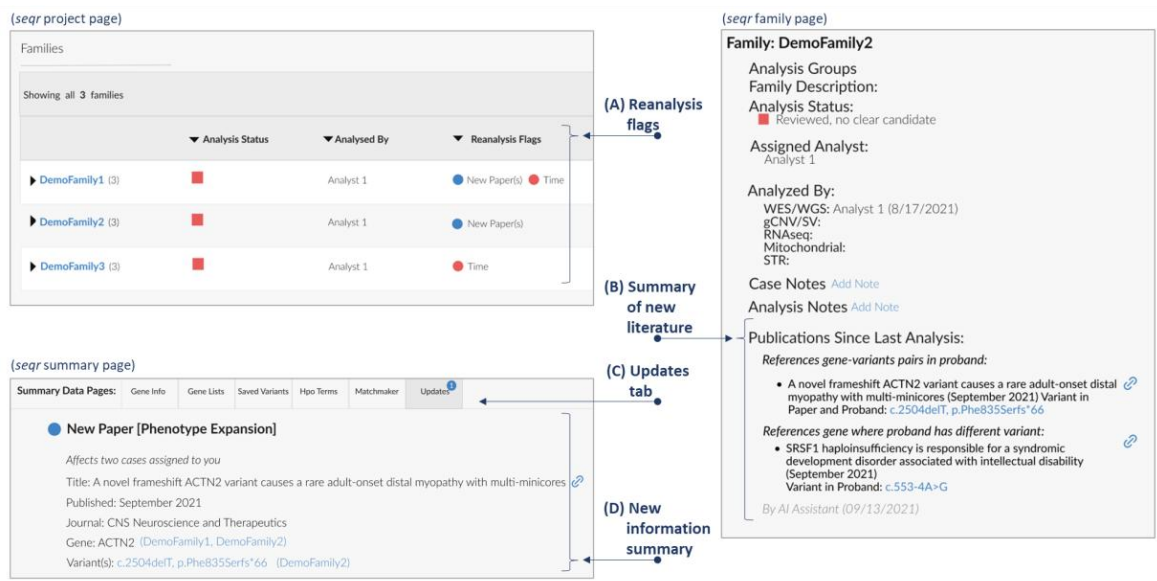

Figure 5: Prototyped feature of an AI assistant flagging cases for reanalysis.
Prototype designed for display in *seqr* platform family and summary data pages. Feature functions: (A) Flags on project page that identify if case is a candidate for reanalysis based on elapsed time or relevant new paper. (B) AI-assistant generated summary of publications since last analysis on family page. (C) Tab in summary page to flag cases across projects. (D) Information on summary page about new papers and affected cases.

### 6.2 Designing the "Thing Right": Exploring Interactions with an AI Assistant (RQ3)

Drawing on the findings from the group co-design workshop, we developed a prototype of an AI assistant to explore different interactions between genetic professionals and AI assistant during WGS analysis (Figure 5-7). In follow-up design walk-through sessions with variant analysts, we identified two themes in the feedback elicited from the prototype: (1) balancing comprehensive and selective evidence and (2) interpreting and verifying information as a collaborative effort.

*6.2.1 Design of AI Assistant Prototype.* We created our prototype of an AI assistant within the existing sequencing analysis tool (*seqr*) used by our primary research site. To populate the information shown in the prototype, we used information from actual publications about a gene (*ACTN2*) [14, 34, 51, 74, 75]. The AI assistant performs three tasks: (1) flagging cases for reanalysis, (2) aggregating and synthesizing key information about a gene and variant, and (3) drafting presentations for variants of interest. We selected these tasks as they were brainstormed by multiple participants in the group workshop and included in their sketches. Below we further describe how participants envisioned interacting with the AI assistant in their sketches and how this informed the prototype design.

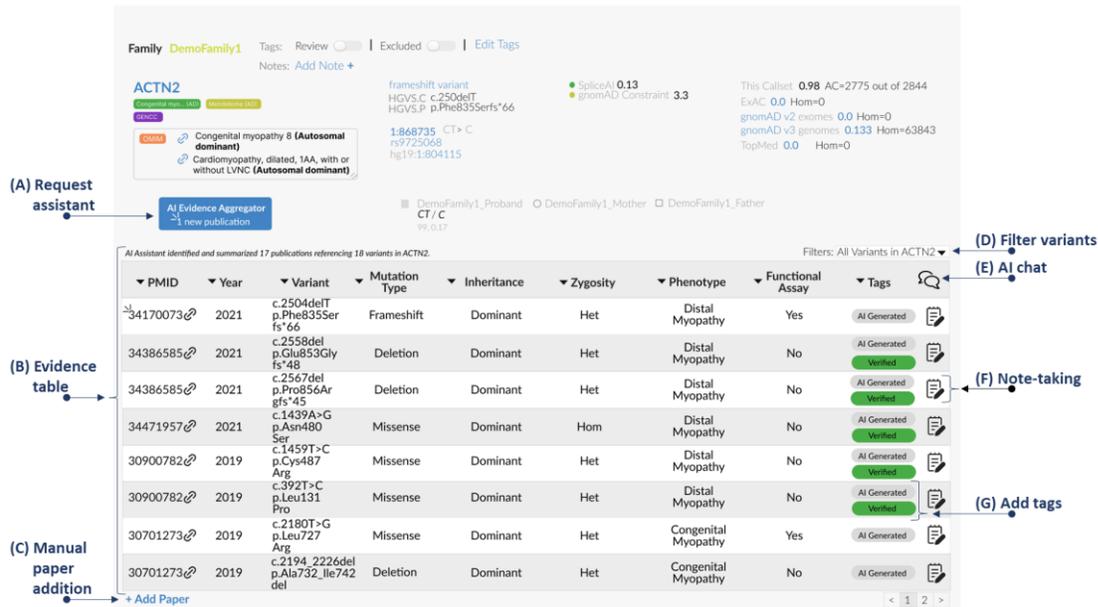

Figure 6: Prototyped feature of an AI assistant aggregating and synthesizing publication evidence about a gene and its variants. Prototype designed for display in *seqr* platform variant search results. Feature functions: (A) Button to request AI Assistant to aggregate evidence about variant and display in table form below. (B) AI-assistant generated evidence table of observed variants synthesized from aggregated relevant publications. (C) Option for analyst to manually add papers missed by AI Assistant. (D) Toggle option to filter table for all variants in gene or only the variant observed in the patient. (E) AI-assistant chat initialization. (F) Take notes about a variant / paper in corresponding table row. (G) Add tags to signify that information was AI-generated, that information from AI Assistant has been verified, or that the row should be excluded (not pictured).

*Task 1: Flagging cases for reanalysis.* One participant [VA#1] sketched the AI assistant flagging cases for reanalysis within the project page of *seqr* that lists all cases (individuals/families sequenced as part of a project) (see Appendix A.1). In their sketch, the user could select to view cases that matched different reanalysis triggers (e.g., all cases that were candidates for reanalysis based on elapsed time or new scientific findings). In our prototype, cases were flagged as candidates for reanalysis within that project page (Figure 5.A). The user could then navigate to a page with details about the individual and family in the case. To help the user evaluate if they should reanalyze the case, this page would contain a new section generated by the AI assistant with information about new publications since the last analysis (Figure 5.B). Our prototype also included a new "Updates" tab within the summary data section (a tabular webpage with information about all projects) of *seqr* (Figure 5.C). This tab aims to help users understand how new publications may affect cases across projects. The Updates tab had a feed generated by the AI assistant with information about new papers and unsolved cases that might be affected (Figure 5.D).

*Task 2: Aggregating and synthesizing key information about a gene and variant (during initial analysis or reanalysis of WGS data).* Participants sketched three mechanisms of interacting with an AI assistant for aggregating and summarizing information about a gene and variant (see Appendix A.2). In one sketch, the AI assistant generates a table with key information from publications about a gene and variant. In another sketch, the AI assistant generates notes with information about the variant. In the third sketch, the user can ask the AI assistant to retrieve and synthesize literature on a particular topic (e.g., "pull literature on GRHL3 [a gene] and 'X' disease"). We combined these three types of interaction mechanisms

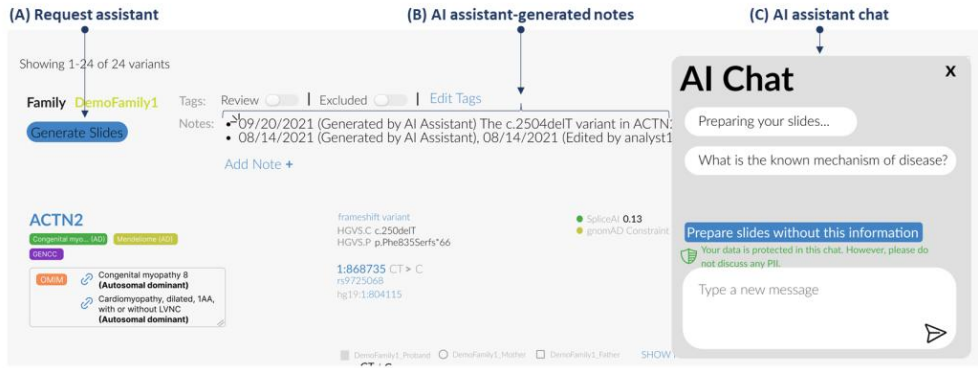

Figure 7: Prototyped feature of an AI assistant generating presentations for team meetings. Prototype designed for display in *seqr* platform variant search results. Feature functions: (A) Button to request AI assistant drafts presentation slides. (B) AI assistant-generated notes which may be used by AI assistant for drafting presentation. (C) Upon request to draft presentation slides, AI assistant chat would be initiated with a pop-up window. AI assistant would ask questions to probe the user for missing information based on a preset presentation template.

in a human-in-the-loop feature within our prototype (Figure 6). In the list of gene-variant pairs for an individual returned by *seqr*, each gene-variant pair section has a button to request evidence aggregation from the AI assistant (Figure 6.A). The AI assistant then generates a table with information from scientific papers about observed variants within the gene (Figure 6.B). In addition to filtering on any column in the table, the prototype also includes preset filters to quickly switch between viewing all observed variants in the gene or viewing only the specific variant observed in the individual (Figure 6.D). Filtering can support users in sensemaking by reducing information overload [32, 81]. The user can also take notes about a row in the table, with the AI assistant generating a modifiable note draft as a starting point (Figure 6.F). We included several tools to help users manage "imperfect" content from the AI assistant and provide feedback from the user to the system [9, 11]. One feature allows users to manually add papers that were not included by the AI assistant (Figure 6.C). Users can also directly edit fields in the table to correct AI-generated information errors. The generated table also has a column to add tags, allowing users to easily convey if they have verified the information in a row or if they think a row should be excluded from the table (Figure 6.G). In addition to supporting the user in managing imperfect content from the generative AI model, these refinement tools collect user feedback that can be leveraged to train and improve the model. Finally, the user can also chat with the AI assistant to ask follow-up questions after viewing the table (Figure 6.E).

*Task 3: Drafting presentations for variants of interest (during initial analysis or reanalysis of WGS data).* One participant [VA#8] sketched the AI assistant generating materials for presentations (see Appendix A.3). The AI assistant would extract content from *seqr* information fields and would ask the analyst questions about missing information needed for the presentation based on a preset presentation template. Drawing from their sketch for our prototype, we added a human-in-the-loop feature to the variant screen where the user could request that the AI assistant generate slides based on selected notes about a variant (Figure 7). The AI assistant would generate the slides, asking questions to probe the user for information to include within the presentation.

6.2.2 *Themes in Feedback Elicited by Prototype.* To elicit feedback from analysts on the design of the AI assistant, we presented the prototype in follow-up design walk-through sessions. We identified two themes in participant feedback about the prototype: (1) balancing comprehensive and selective evidence and (2) interpreting and verifying information as a collaborative effort.

*Balancing comprehensive and selective evidence.* Participant reactions to the features for flagging cases for reanalysis (Figure 5) and aggregating and synthesizing information from publications (Figure 6) highlighted the need to consider the appropriate balance between comprehensive and selective evidence in these features. Participants often asked about the criteria used to determine when a case is flagged for reanalysis because of new papers, worried about *"too much noise"* [VA#4, VA#8] if the criteria were not selective enough. In contrast, participants preferred viewing more comprehensive evidence in the feature to aggregate and synthesize information from publications about a gene and variants. In the first iteration of the prototype, the AI assistant filtered papers presented in the table to match the individual's phenotype. However, participants had negative reactions to this selective filtering by the AI assistant. One participant [VA#5] found it hard to form a mental model of the complexities of a gene with some papers and variants filtered out. Another participant [VA#4] did not trust the AI assistant to appropriately match the individual's phenotype with the phenotypes from the papers. Most participants preferred to see all information at first and then apply their own filters.

*Interpreting, editing, and verifying AI-generated information as a collaborative effort.* Participants envisioned collaboratively interpreting and verifying information aggregated and synthesized about a gene and variant. Because individuals in different cases may have variants in the same gene, participants often highlighted that the AI-generated table could be shared between analysts in their institution or even with those in other institutions. Participants envisioned reusing the work of others by viewing (1) if the row in the table had been verified by another user, (2) the edits made to the information in the table, and (3) notes taken by another user. Sharing these artifacts could reduce duplicated work and help analysts better interpret the information. Most participants appreciated the ability to view if a row had been verified by another analyst, noting that they may not dig further into the paper in a verified row unless it was highly applicable to their case. One participant, however, raised concerns that verifying information in the table would be too much *"additional work"* [VA#4]. Participants also discussed the level of edit access that should be given to others, noting the importance of being able to understand who edited the table and why they corrected the information generated by the AI assistant. One participant [VA#4] thought that analysts in training or outside their institution should be granted comment-only access, while other analysts in the institution could have edit access because they would *"use it responsibly"* [VA#4]. Another participant [VA#8] raised concerns that edits from others would mask the inaccuracies of the AI assistant, making it harder to calibrate their trust in the AI assistant. Participants also appreciated the ability to view notes taken by other analysts because the notes could help them understand the information faster while reducing duplicated work. To calibrate their trust in the note, participants wanted the ability to view who wrote the note (e.g., the AI assistant or another analyst). One participant [VA#2] explained how they might use information about the notetaker (e.g., their experience level) to determine their level of trust in the note. Participants appreciated the ability to view the notes of others but were concerned that there would be confusion about where to write different types of notes. They also felt that it might be overwhelming to identify the notes that could help them interpret their case. Currently, analysts often record information about publications in the notes section for a particular variant in the case. Because some notes may apply to other cases, one participant had experimented with concatenating all notes into a *"master knowledge set."* However, they found this concatenation unhelpful because the notes were *"all over the place"* [VA#5] with a mix of repeated and case-specific information.

## 7 DISCUSSION

In this study, we examined how generative AI could be used to support knowledge work, focusing on the use of LLM-based generative AI to aid genetic professionals in WGS analyses. Our findings from both study phases highlight that individual and collaborative sensemaking is a challenging aspect of WGS analysis that genetic professionals envision and prioritize for AI support. Many of the current challenges in WGS analysis involve making sense of the growing information about genes and variants. In the needs elicitation phase of the study, participants highlighted issues synthesizing information about genes and variants from different information sources, sharing findings with other genetic professionals, and identifying previously unsolved cases that may be affected by new scientific findings. These challenges were reflected in the two tasks of an AI assistant prioritized by participants during the design ideation phase: (1) flagging cases for reanalysis based on new scientific findings and (2) aggregating and synthesizing key information about genes and variants. Next, we discuss how genetic professionals prioritized these tasks for AI support and the ways in which the AI tool envisioned by our participants differs from the AI tools currently being developed. We also describe opportunities for AI to augment the sensemaking process of genetic professionals, focusing on the two sensemaking tasks prioritized by participants. Finally, we propose design considerations for supporting individual and collaborative sensemaking with generative AI.

## 7.1 Determining Tasks of an AI Assistant: Matching AI Capabilities with User Needs

To mitigate challenges in determining the appropriate tasks of an AI system, we approached our co-design sessions from a blended user-centered and technology-centered approach [5, 93, 96]. Because most participants had a limited knowledge of generative AI before the study, we began the group co-design workshop by discussing the capabilities and limitations of generative AI. During this workshop, several participants discussed how they had briefly experimented with LLM-powered tools (e.g., ChatGPT and Bard), asking these tools to perform tasks related to WGS analysis. Although the tools performed some aspects of tasks well (e.g., explaining disease mechanisms), participants also noticed that the tools made mistakes (e.g., providing false paper references). We observed that the blended user-centered and technology-centered approach from the group workshop was reflected in how participants prioritized tasks for AI support in the follow design walk-through sessions. In these sessions, participants often prioritized effort-intensive tasks that they thought could be performed well by AI.

Following a blended user-centered and technology-centered approach highlighted a discrepancy between the types of AI support envisioned by genetic professionals and the AI tools currently being developed for WGS analysis. In the prototype designed with genetic professionals, an AI assistant performs select tasks (e.g., flagging cases for reanalysis, aggregating and synthesizing information, and drafting presentations), with a human-in-the-loop level of delegation [52]. However, the WGS analysis process remains largely driven by genetic analysts. In contrast, the AI tools currently being developed from a purely technology-centered approach automatically analyze or reanalyze entire cases, with little discussion of how genetic professionals may interact with these tools or interpret their outputs [4, 19, 45, 50, 57]. This discrepancy between the AI tools envisioned by genetic professionals and the AI tools being developed could lead to future issues with tool adoption and workflow integration. Some participants also worried about the impacts that generative AI could have on their job security and expressed concerns that they may be designing their replacements. These concerns are especially noteworthy given the level of expertise and domain knowledge required to conduct WGS analysis.

## 7.2 Sensemaking in WGS Analysis: Current Process and Opportunities for AI Support

Our participants prioritized using generative AI to support their sensemaking when reviewing and interpreting variants in a case and selecting cases for reanalysis. Next, we further discuss the current sensemaking processes of genetic professionals, the opportunities for AI support, and the potential changes that may occur to sensemaking with this support.

*7.2.1 Reviewing and Interpreting Variants in a Case.* During the initial analysis and reanalysis of an individual's WGS data, analysts manually review and interpret tens to hundreds of gene-variant pairs [2]. Currently, analysts conduct most of the foraging and sensemaking activities alone (Figure 8). In the foraging activity, analysts search for different types of evidence about gene-variant pairs, synthesizing the relevant information. In the sensemaking activity, analysts build a model of information known about the gene and variant, determine how the individual's variant relates to this model, and share their findings. Analysts may be reviewing prior cases (during reanalysis), genes that they have seen before in other cases, or genes that have appeared in other analysts' cases. However, duplicated work often occurs because the tools currently used by analysts are limited in supporting sensemaking translucence [28] and distributed sensemaking [25]. These tools require additional work from the analyst, such as taking detailed notes or discussing information with others through data-sharing platforms. Although taking detailed notes could better support the sensemaking process of other users and that of the current user if they return to the case for reanalysis [31], analysts may not have the time or bandwidth for this activity. Explicit sharing of information between users can improve sensemaking but this practice can also increase the user's workload and negatively impact their perceptions of tools [20].

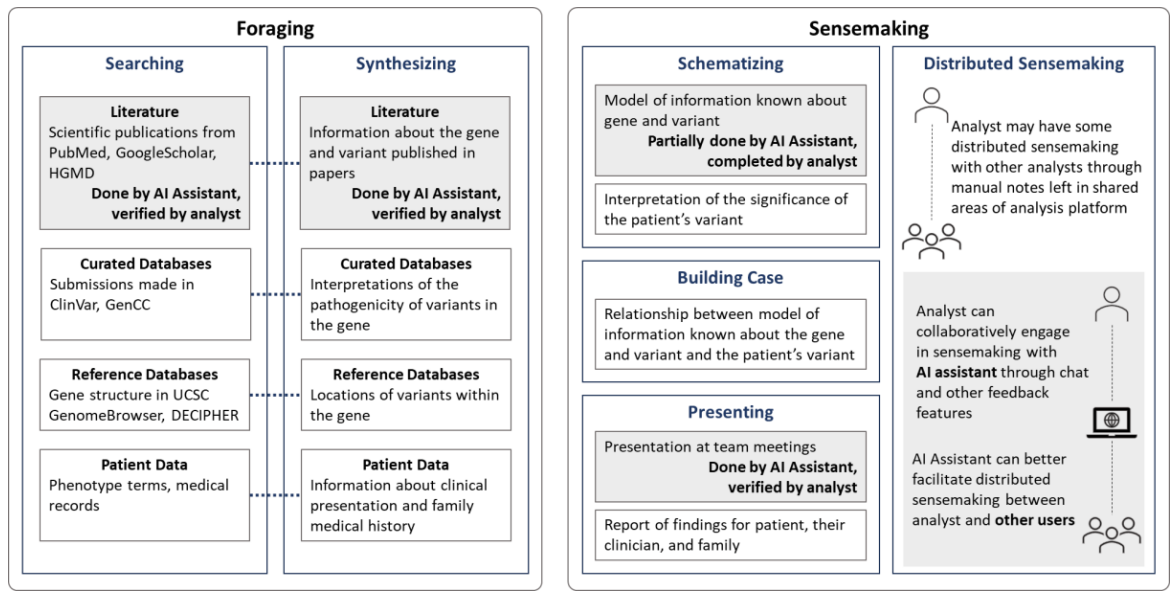

Figure 8: Sensemaking process when interpreting variants with the introduction of prototype AI assistant. Gray boxes represent sensemaking activities which are currently performed by an analyst but are human-in-the-loop processes with involvement of our prototype AI assistant. Non-gray boxes represent activities reserved for analyst completion without assistance by our AI assistant prototype. Within the foraging searching and synthesizing processes, examples of data sources and data types for each, respectively, are connected by dotted lines.

Our participants envisioned and designed an AI assistant that would support the sensemaking process of an individual analyst and better facilitate distributed sensemaking between analysts. In the foraging activity, the AI assistant could search for relevant scientific papers and synthesize information about the gene and variant. In the sensemaking activity, the AI assistant could support the analyst by partially schematizing the information by building a table and drafting presentations for variants of interest. When new evidence is discovered (e.g., a new paper is published about a gene), the AI assistant could augment and highlight this evidence in the artifact to support reanalysis. Participants envisioned the AI assistant creating evolving artifacts that could be shared between cases, analysts, and even institutions to support distributed sensemaking and verification of AI-generated content. Because the AI assistant is generating sensemaking artifacts, sensemaking translucence and distributed sensemaking may be supported without much additional work for analysts.

Future research will need to evaluate how this feature of an AI assistant affects analysts' sensemaking processes. Instead of doing all sensemaking steps themselves, analysts will be interpreting and verifying information synthesized by the AI assistant. We expect that these changes to the sensemaking process will reduce the time and workload of analysts while improving their sensemaking of variants. However, two concerns could negatively impact this expectation. First, people often decide not to use others' artifacts in their sensemaking because of the work required to verify the artifacts [48]. If verifying the artifacts generated by the AI assistant is too burdensome, analysts may entirely disregard the artifacts. Second, the analyst's sensemaking process could be negatively impacted when foraging and sensemaking tasks are delegated to an AI assistant instead of being directly performed by the analyst. Automation can reduce a user's workload but high levels of automation can also reduce awareness of information [23]. Prior work, however, has found that users have better task performance when they can spend more time processing results, instead of searching for documents and extracting the relevant information [85]. This finding supports our expectation that automatically aggregating and synthesizing information about genes and variants will improve sensemaking.

*7.2.2 Selecting Cases for Reanalysis.* Sensemaking also occurs when analysts need to identify unsolved cases that should be reanalyzed because of new scientific findings. Over 1,000 papers are published each year with new disease descriptions or new gene-disease links [22]. Key information needs to be synthesized from new papers, incorporated into models of information known about genes and variants, and applied to unsolved cases to hypothesize if any would benefit from reanalysis with the new information. However, analysts are limited in their ability to identify cases that should be reanalyzed because of new scientific findings because of other responsibilities and the number of papers published each year. Participants envisioned and prioritized an AI assistant that reasons about new papers and unsolved cases to identify and flag cases for reanalysis based on new scientific findings. An AI assistant that supports sensemaking by automatically synthesizing new papers and identifying potentially affected cases could increase the diagnostic yield of reanalysis. LLM-based AI's mastery of language could be capable of determining relevant papers by reasoning based on patient phenotype while also having the capability to summarize rationale for flagging a case for reanalysis. However, one challenge will be appropriately alerting analysts to cases that would benefit from reanalysis. If cases are often incorrectly flagged for reanalysis, analysts may start to ignore the AI assistant's suggestion to reanalyze cases. When automated tools can appropriately detect unsolved cases affected by new scientific findings, reanalysis may change from a periodic event to a continuous process. Instead of reanalysis being started by a specific trigger (e.g., a request from the patient's clinician), automated tools could continuously reanalyze cases. Analysts could then reanalyze cases only when the system has detected a high probability of new information leading to a diagnosis. Even if such a system is created, sociotechnical barriers to this type of reanalysis process may exist, such as the funding structures to reimburse this type of continuous reanalysis [2] and analysts' trust in this type of system. Additionally, clinicians, individuals with rare diseases, and their families can currently request reanalysis in some cases. Moving from a periodic, manual reanalysis process to a continuous, more automated process could reduce this agency. Future research can examine how other roles in WGS perceive and may be affected by continuous reanalysis.

## 7.3 Design Considerations for Using Generative AI to Support Sensemaking

Although our study was situated in the context of WGS analysis, the sensemaking work prioritized for AI support by genetic professionals has parallels to the sensemaking work performed by knowledge workers in other domains. Similar to other knowledge workers, genetic professionals must stay abreast of new information, synthesize information from various data sources, and make decisions based on their synthesis [26, 36, 40]. Our participants also highlighted similar challenges in synthesizing information from different sources (without missing key information), documenting the rationale for decisions, and collaborating with others. To address these frequent sensemaking challenges, tools have been developed to aid users in creating artifacts that can support their own sensemaking processes and be shared with others to also support their sensemaking [28, 31, 46–48]. With the increased abilities of generative AI models to interpret and synthesize information, recent work has developed AI systems that can automatically generate interactive artifacts to support sensemaking [26, 37, 40]. However, further research is needed to understand the changes to sensemaking when sensemaking artifacts are created by generative AI models and the design of interactions between users and generative AI models to support sensemaking [37, 81]. We next contribute and discuss three design considerations for using generative AI models to support sensemaking.

1. *Facilitating distributed sensemaking.* While generative AI models can create artifacts that support the sensemaking process of an individual user, these artifacts can also be shared between users to facilitate distributed sensemaking. Similar to prior studies [95, 97], our participants considered explainability, feedback, and trust when envisioning how they would interact with AI-generated content. Sharing AI-generated artifacts between users could reduce the amount of work needed to verify and interpret the artifacts while facilitating distributed sensemaking between users. To reduce the work needed to verify the artifacts [48], users could view if portions of the artifact had been marked as correct by other users and any edits made to the AI-generated information. However, one participant in our study raised concerns that this design could prevent users from appropriately calibrating their trust in the AI by masking the inaccuracies of the generative AI model. Designs that allow for corrections of AI-generated information by other users may need to clearly mark the corrections made to the AI-generated information to support trust calibration. Additionally, providing information about the user who made the corrections could be important as users may need to decide if they trust the correction. To facilitate distributed sensemaking and support users in interpreting AI-generated artifacts, notes taken by others should be visible. Because not all notes may be relevant, generative AI models could even extract and synthesize notes from other users that apply to the context of the current user.

2. *Supporting initial sensemaking and re-sensemaking.* Sensemaking is both affected by time and affects our understanding of time [33]. Information-related tasks may be revisited later to understand the rationale for a previous decision or if there is new information that may change the outcome of the task. The user returning to the task may or may not be the same user who performed the task originally. Viewing AI-generated artifacts from prior instances of information-related tasks can help users understand the rationales behind decisions [47]. Generative AI models could be continuously editing artifacts, creating artifacts that evolve over time. In addition to highlighting new information within the artifact since the last instance of the task, generative AI models could also support sensemaking translucence [28] by helping a user understand their previous sensemaking process. For example, an AI assistant could summarize the notes taken and questions asked by a user during prior instances of performing a specific information-related task.

3. *Combining evidence from multiple modalities.* In information-related tasks, users may need to combine information from multiple modalities during sensemaking. For example, search and rescue workers receive, aggregate, and synthesize information from multiple modalities, such as text, pictures, and GPS locations [38]. Recent developments have led to increasing numbers of multimodal generative AI models that can accept and synthesize multiple input modalities [77]. Such models could support sensemaking by synthesizing information from different modalities in one view. For example, a participant in our study sketched the generative AI model combining text from scientific publications and a visual depiction of the gene structure to create an image that compares the location of the individual's variant in the gene with the locations of variants from published papers.

### 7.4 Study Limitations

Our study had two main limitations. First, although the needs elicitation phase of the study had participants from multiple institutions, all participants in the design ideation phase of the study were from the same institution. Other institutions may have different policies, practices, and tools that could impact how participants envision the design of an AI assistant. However, the features of an AI assistant developed in the design ideation phase aimed to address challenges highlighted by the first-phase participants across research sites, providing evidence that the findings may apply across institutions. Second, participation in both phases of the study was voluntary. As a result, participants with greater openness towards AI may have been more motivated to participate. Further research is needed to fully gauge the willingness of genetic professionals to engage with AI-based tools.

## 8 CONCLUSION AND FUTURE WORK

In this research, we examined how generative AI could be used and designed to support domain experts in performing knowledge work. We focused on designing a generative AI-based assistant to aid genetic professionals in WGS analysis, enabling them to find diagnoses more quickly for rare genetic disease patients. In the first phase of our study (needs elicitation), we identified the current challenges faced during WGS analysis. In the second phase of our study (design ideation), we developed a prototype of an AI assistant to support WGS analysis. The AI assistant aims to increase diagnostic yield and reduce time to diagnosis by (1) flagging cases for reanalysis and (2) aggregating and synthesizing key information about genes and variants. As our main contributions to the HCI community, we contribute a detailed empirical study of WGS analysis, an understanding of how domain experts envision generative AI supporting their knowledge work, and three design considerations for supporting sensemaking with generative AI. Although we collected feedback about the prototype, further research is needed to understand how analysts interact with an AI assistant during their analysis of WGS data. In our future work, we plan to conduct task-based user testing with genetic professionals to evaluate how the AI assistant impacts their sensemaking and workflow.

# A APPENDICES

## A.1 Participant sketch of an AI assistant flagging cases for reanalysis (created by VA#1).

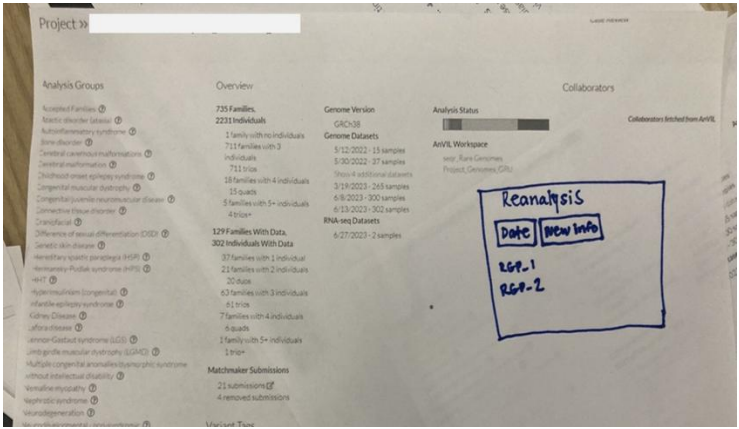

## A.2 Participant sketches of an AI assistant aggregating and synthesizing information about a gene and variant.

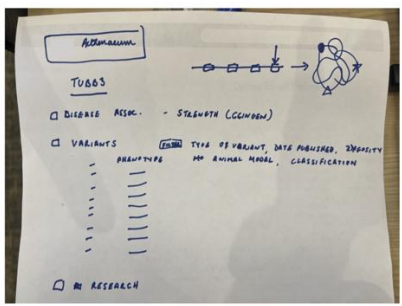

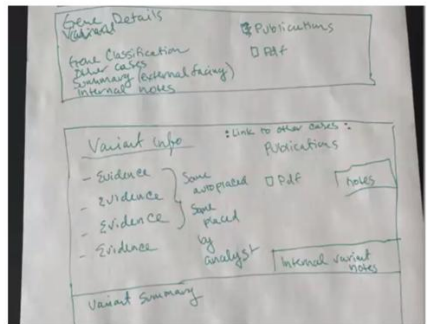

Sketched by VA#2

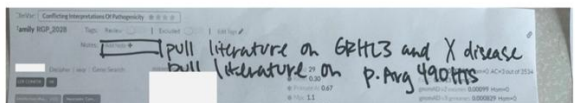

Sketched by VA#4

Sketched by VA#5

## A.3 Participant sketch of an AI assistant generating a presentation about a variant of interest (created by VA#8).

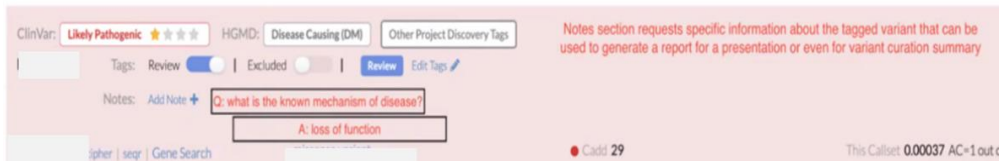